\documentclass[runningheads]{llncs}
\usepackage{graphicx}
\usepackage{changes}
\usepackage[T1]{fontenc}
\usepackage{soul}
\usepackage{url} 
\usepackage{xurl}
\begin{document}
\title{From Text to Knowledge with Graphs: modelling, querying and exploiting textual content\thanks{This discussion was organised by the actions DOING and ROCED supported by the CNRS GDR MADICS of the French government.}}
%
%
\author{
Genoveva Vargas-Solar \inst{1}
\and
Mirian Halfeld Ferrari Alves\inst{2}
\and
Anne-Lyse Minard Forst\inst{3}
}
%
%
\institute{
CNRS, Univ Lyon, INSA Lyon, UCBL, LIRIS, UMR5205, 69622, Villeurbanne \\
\email{genoveva.vargas-solar@cnrs.fr}
\and
University of Orléans, LIFO, Orléans \\
\email{mirian@univ-orleans.fr}
\and
University of Orléans, LLL, Orléans \\
\email{anne-lyse.minard@univ-orleans.fr}
}

\maketitle              
\begin{abstract}
This paper highlights the challenges, current trends and open issues related to the representation, querying and analytics of content extracted from texts.  
The internet contains vast text-based information on various subjects, including commercial documents, medical records, scientific experiments, engineering tests, and events that impact urban and natural environments. Extracting knowledge from this text involves understanding the nuances of natural language and accurately representing the content without losing information. This allows knowledge to be extracted, inferred, or discovered. To achieve this, combining results from various fields, such as linguistics, natural language processing, knowledge representation, data storage, querying, and analytics, is necessary. The vision in this paper is that graphs can be a well-suited text content representation once annotated and the right querying and analytics techniques are applied. This paper discusses this hypothesis from the perspective of linguistics, natural language processing, graph models and databases and artificial intelligence provided by the panellists of the DOING session in the MADICS Symposium 2022 \footnote{The paper summarises the discussion in the panel of the national action DOING in the national symposium MADICS 2022 in Lyon.}.

\end{abstract}
%
%
%
\section{Context}
Text analysis refers to the processing of texts to extract machine-readable facts. It aims to create structured data from free textual content. The process can be seen as cutting up pieces of unstructured and heterogeneous documents into manageable and interpretable data elements. From a computational point of view, this process requires techniques ranging from semi-automatic language processing to decode the ambiguity of human language, knowledge representation, and detection of patterns and trends representing knowledge conveyed in texts (querying in the broad sense). 

The content of texts defines both a syntactic mesh and a concept mesh that can be structured as graphs. Querying these contents can be done at several levels according to the degree of abstraction this graph represents, ranging from simple guidance on the raw text to constructing a database (graph) respecting a certain number of structural constraints.

Depending on the characteristics of the graphs used, it is possible to query the content of texts in different ways:
Information retrieval offers  keyword methods.
\begin{itemize}
    \item Querying graphs via query languages allows  to find structural patterns and calculate aggregations.
    \item Machine learning and data mining methods allow to discover patterns.
    \item Artificial intelligence makes it possible to discover new semantic links between concepts.
\end{itemize}

Text analysis and knowledge extraction are thus addressed from different points of view, perhaps complementary, depending on the exploitation objectives through applications. 
This paper discusses:\\
    \noindent
    [i] How text content is extracted and represented by different data models, such as matrices with term frequencies, and ontologies, by designing graph databases.\\
     \noindent
    [ii] The implications of the modelling choice on the possibilities of querying, updating and knowledge discovery.\\
%
%
%
\noindent
The discussion is organised around a leitmotif question:
\begin{quote}
    {\em How can natural language processing, the Semantic Web, DB design for Graph DBMSs, and machine learning/data mining contribute to model/query text content?}
\end{quote}
This question has been specialised into the following research questions that have guided the analysis, statements and discussions proposed in the paper.
\begin{enumerate}
    \item [RQ$_1$]
    What are the barriers and challenges of textual content processing? What are representative applications? Which experiments  have been representative or  have posed particular challenges?
    
    \item [RQ$_2$]	How to deal with missing information, i.e., incomplete, poor quality or contradictory annotations on property graphs and/or ontologies representing text content? What are the impacts on querying and maintenance? What issues need to be addressed?
    
    \item [RQ$_3$] Maintenance and updating of text content representations in the form of property graphs or ontologies:
    \begin{itemize}
        \item How to make annotations evolve according to the evolution of repositories, ontologies and corpora?
        \item How to validate new knowledge deduced, discovered or explicitly inserted?
    \end{itemize}
\item  [RQ$_4$] What are the possibilities of querying the different representations of the text content?
\begin{itemize}
    \item [RQ$_4.1$] How are they exploited in NLP-based applications?
    
    \item [RQ$_4.2$] How can  querying in the "analytics" sense (data science queries with community discovery, centrality and link discovery operations) on graphs be addressed?
    
    \item [RQ$_4.3$] What are the solutions proposed by graph-based DBMSs? How can they be positioned concerning solutions such as ML pipelines/DS pipelines?
    
    \item [RQ$_4.4$] Is it relevant to think of a SPARQL-extended query language where the user can include a 'function' concerning graph data analysis? What are the advantages? What uses?  What challenges?
\end{itemize}
\end{enumerate}

The remainder of the paper proposes answers to these questions from perspectives of different disciplines: Natural Language Processing, Semantic Web, Databases and Machine learning. Existing work and open issues are identified. They all provide complementary solutions and open research perspectives to fully address the challenge of transforming textual content into knowledge to answer different questions.

Accordingly, the paper is organised as follows. Section \ref{sec:NLP} elaborates on the problems, existing solutions and open issues on how to represent textual content by exploring its syntax and semantics. 
 Section \ref{sec:websem} addresses text content representation challenges from the semantics point of view. It discusses how the semantic Web can contribute to representing knowledge from textual content and how it is maintained, queried and inferred. 
Section \ref{sec:dbgraphs} discusses how graph database management systems can contribute to storing text content and can provide efficient methods to maintain, query and complete knowledge. Section \ref{sec:ml} gives perspectives on how machine learning and artificial intelligence play a role in discovering knowledge within text content and from graphs when used for modelling it. Section \ref{sec:conclusion} concludes the paper and wraps up answers to the questions we used to discuss about text to knowledge transformation.

\section{Extracting knowledge from text: NLP problems and open issues}\label{sec:NLP}
The goal of Natural Language Processing (NLP) is to allow a computer to "understand" the textual content, including the contextual nuances of the language within them \cite{nadkarni2011natural}.   NLP techniques can accurately extract text information and insights to categorise and organise the associated documents.  Language-processing systems \cite{idnay2022systematic} have been designed by symbolic methods based on hand-coding rules and a dictionary lookup, such as writing grammar or devising heuristic rules for stemming.

Since the 90s, natural language processing research has relied on machine learning that calls for using statistical inference to learn linguistic rules through the analysis of large corpora\footnote{Corpora refers to the plural form of a corpus, which is a set of documents, possibly with human or computer annotations.} automatically. 
\paragraph{Natural Language Processing Tasks}
 can be classified according to their purpose: 
\begin{itemize}
\item {\em Text and speech processing (optical character recognition, speech recognition, speech segmentation, text-to-speech, word segmentation)}: the general principle is that a text in a document of some format (image, sound clip, text) determines the textual representation of the content, which can be the words/sentences composing textual content. The objective can be more or less challenging depending on the characteristics of the language in which the textual content has been produced.

\item  {\em Morphological analysis (lemmatisation, morphological segmentation, part-of-speech tagging, stemming)}: the objective is to separate the lemmas or morphemes of the words of a text to produce normalised representations or to identify parts of speech (POS) within sentences (e.g., noun, verb, adjective). 

\item Syntactic analysis (grammar induction, sentence breaking, parsing) seeking to understand the structure of natural languages. The principle is to generate a grammar that describes a language's syntax or find the sentence boundaries within chunks of texts identifying punctuation marks and building parse trees (grammatical analysis) of a given sentence that represent relationships (dependencies) between words with (dependency parsing) or without probabilities (constituency parsing that produces probabilistic context-free  or stochastic grammars).

\item  {\em Lexical semantics of individual words in context (lexical semantics, distributional semantics, named entity recognition, sentiment analysis, terminology extraction, word-sentence disambiguation, entity linking)}: the objective is  to determine the computational meaning of individual words in context or from data. Methods are intended to 
(i) identify from a stream of text which items   map to proper names (e.g.,  people or places), and determine the type of such name  (e.g. person, location, organisation); (ii) extract subjective information (e.g., sentiments); (iii) identify terminology; (iv) disambiguate words with more than one meaning.

\item {\em Relational semantics (relationship extraction, semantic parsing, semantic role labelling)}:  identify the relationships among named entities, identify and disambiguate semantic predicates and semantic roles, and produce formal representations of the semantics of a piece of text.

\item {\em Discourse analysis goes beyond the semantics of individual sentences (conference resolution, discourse analysis, implicit semantic role labelling, recognising textual entailment, topic segmentation and recognition, argument mining)}. Given a sentence or a more significant chunk of text, determine which words ("mentions") refer to the same objects ("entities"). The more general task of co-reference resolution includes identifying  "bridging relationships" involving referring expressions, discourse parsing for determining the nature of the discourse relationships between two sentences, recognising textual entailing, identifying topics within text segments, and identifying argumentative structures.

\end{itemize}

\paragraph{Textual content modelling.}
It is essential for NLP methods to efficiently represent syntax and semantics and exploit them to implement applications like question and answering, automatic summarisation, grammatical error correction, and converting information from computer databases or semantic intents into readable human language, among others. The literature\cite{eisenstein2019introduction,maulud2021state}  recognises that NLP graphs can represent 
the co-occurrence of words in documents, knowledge graphs, and sentences as graphs. For every type of graph, edges can have different semantics. For example, in an ontology, nodes represent concepts and edges semantic relations; in a co-occurrence, graph nodes represent words and edges absolute or relative frequencies between two connected words or n-word window intervals, etc.

From a knowledge representation point of view, the question is how to transform textual content into graphs that fully represent the content.
From a database point of view, it is a question of knowing which graph models are the most suitable for designing these graphs, e.g. property graphs, DAGs (Direct Acyclic Graphs), unDAGs (Undirected Acyclic Graphs), etc. Sections \ref{sec:ml} and \ref{sec:dbgraphs} will further discuss these aspects.

\paragraph{Text mining and NLP techniques}
  are powerful tools for extracting structured semantic information from unstructured machine-readable text. These techniques can help uncover insights contained within text in a scalable and efficient manner by identifying and analysing explicit concepts and relationships. Some common text mining and NLP techniques include structure extraction, tokenisation, acronym normalisation, lemmatisation, decompounding, and identifying language, sentences, entities, relations, phrases, and paragraphs. These techniques help to make sense of large amounts of unstructured text data and extract valuable insights from it. As the amount and rate of text data increases, these techniques have become increasingly crucial for analysing unstructured data. Using text mining and NLP techniques, researchers can gain a deeper understanding of the world around us and make more informed decisions based on the insights extracted from large amounts of natural text data.

Many different classes of machine-learning algorithms can be applied to NLP tasks. Algorithms receive as input a large set of "features" that are generated from the input data that result from preprocessing texts. Machine learning algorithms, such as decision trees, are the basis of systems of complex if-then rules similar to existing hand-written rules. Part-of-speech tagging allows the use of hidden Markov models. 
 NLP research has also used neural networks. 
 
\paragraph{Current trends and open issues.}
 In the field of NLP, there is a growing interest in exploring the “cognitive” aspects of natural language, as well as multilingualism and multimodality. Additionally, there are trends towards moving away from symbolic representations and evolving from rule-based to supervised, weakly supervised, and representation learning methods, as well as end-to-end systems \cite{jozefowicz2016exploring}.

{\em [RQ$_1$] What are the possibilities of querying the different representations of the text content? How are they exploited in NLP-based applications?} 
 
 In this discussion, we are interested in the expression power of graph-based representations of the syntaxis and semantics textual content that can then be processed to extract explicit and implicit knowledge. The focus is also on different types of querying approaches that can be applied on top of these textual content representations. For example, queries searching for factual content related to the probabilistic or effective presence of terms in textual documents as in information retrieval, aggregation queries like which is the most recurrent drogue referred to in a collection of clinical cases;  inferring the protocol for medical follow-up for the treatment of disease within clinic cases described in documents; correlating a specific pattern of symptoms and the use of drogues that can deal to the observation of collateral effects reported by physicians in clinical cases.

\begin{enumerate}
    \item 
    {\em [RQ$_2$] What are the barriers and challenges of textual content processing for NLP methods? What are representative applications? Which experiments  have been representative or  have posed particular challenges?}

   NLP has made significant progress in developing methods for representing and analyzing human language computationally. However, NLP methods still face several barriers and challenges when processing textual content:

\begin{enumerate}
    \item Contextual words and phrases and homonyms: Words can have different meanings depending on the context in which they are used. This can make it difficult for NLP methods to accurately interpret the intended meaning of a word or phrase.

    \item Synonyms: Different words can have the same meaning, which can make it difficult for NLP methods to accurately identify the intended meaning of a word or phrase.

    \item Irony and sarcasm: These can be difficult for NLP methods to detect and interpret accurately.
\item Ambiguity: Textual content can often be ambiguous, making it difficult for NLP methods to accurately interpret the intended meaning.
\item Errors in text or speech: Errors in text or speech can make it difficult for NLP methods to accurately process and interpret the content.
\item  Colloquialisms and slang: These can vary greatly between different regions and cultures, making it difficult for NLP methods to accurately interpret their meaning.
\item Domain-specific language: Different domains can have their own specific language and terminology, which can make it difficult for NLP methods to accurately process and interpret the content.
\end{enumerate}

In addition to the challenges faced by NLP methods, there is also the issue of bias. Unsupervised AI models that automatically discover hidden patterns in natural language datasets can capture linguistic regularities that reflect human biases, such as racism, sexism, and ableism \footnote{\url{https://www.analyticssteps.com/blogs/artificial-intelligence-healthcare-applications-and-threats}}. When word embeddings are used in NLP, they can propagate bias to downstream applications, leading to biased decisions. This significant challenge must be addressed to ensure fairness and equity in using NLP methods.

    
    \item [RQ$_3$]{\em  Maintenance and updating of text content representations:}
       [RQ$_3.1$] {\em How to make annotations evolve according to the evolution of repositories?}
Maintaining and updating  text content representations and annotations is a complex process that requires a combination of human expertise and computational methods. It is an ongoing challenge for NLP research, but it is essential for ensuring the accuracy and usefulness of NLP methods. 

\item [RQ$_4$] {\em What are the possibilities of querying the different representations of the text content?}
    %
    [RQ$_4.1$] {\em How can  querying in the "analytics" sense (data science queries with community discovery, centrality and link discovery operations) on graphs be addressed?}
    
    

   Different representations of text content in NLP can be exploited in various applications to improve their performance and accuracy.
    For example, querying text content profiting from vectorial representations (i.e., embeddings representing words as vectors in a high-dimensional space) that capture the semantic relationships between pairs of words can allow more accurate querying of text content. Machine learning algorithms and other computational methods are used to automatically identify relevant information in text and suggest revisions or updates to improve the accuracy and consistency of the content. In NLP-based applications, these representations of text content can be exploited to improve performance and accuracy. For example, in text classification, machine learning algorithms can use word embeddings or other representations of text content to accurately classify documents into predefined categories \cite{ma2018query}. Similarly, in information extraction, NLP methods can use these representations to identify and extract relevant information from unstructured text accurately \cite{kaddari2020natural,khurana2023natural}.

NLP has many applications in healthcare, including improving clinical documentation, supporting clinical decision-making, and improving patient safety. Here are some specific applications of NLP in healthcare \footnote{\url{https://www.mckinsey.com/industries/healthcare/our-insights/natural-language-processing-in-healthcare}}:

\begin{itemize}
    \item Clinical documentation: NLP can be used to extract information from unstructured clinical notes and convert it into structured data that can be easily analysed and used to improve patient care.
    \item Predictive analytics: NLP can be used to analyse electronic medical records and identify patients at greater risk of health disparities, providing an additional level of surveillance.
    \item Clinical decision-making: NLP can be used to support clinical decision-making by providing clinicians with relevant information from a patient’s medical history, as well as the latest research and guidelines.

    \item Patient experience: NLP can be used to improve the patient experience by providing personalised information and support throughout the patient’s journey of receiving care.
\end{itemize}

\end{enumerate}

There are several challenges associated with implementing NLP in healthcare. Some of these challenges include \cite{carriere2021case}:
\begin{itemize}
    \item Variation in language: There are many different dialects and variations of language used in medical records, which can make it difficult for NLP algorithms to accurately understand and interpret the content.

    \item Data standardisation: Poor standardisation of data elements, insufficient data governance policies, and variation in the design and programming of electronic health records (EHRs) can make it challenging to use NLP to fill in the gaps of structured data.

    \item Domain-specific language: The application of NLP methodologies for domain-specific languages, such as biomedical text, can be challenging due to the complexity and specificity of the language used.
\end{itemize}

\section{Modelling knowledge from textual content: Semantic Web}\label{sec:websem}

The Semantic Web is an extension of the World Wide Web that aims to make the web more intelligent and intuitive by enabling machines to understand the meaning of information on the web. One of the main challenges in achieving this goal is dealing with textual content, which is often unstructured and difficult for machines to understand.
To overcome this challenge, text mining techniques can automatically extract meaningful information and semantics from text. Another approach to dealing with textual content in the Semantic Web is the use of knowledge bases such as Wikidata \footnote{\url{https://www.wikidata.org/wiki/Wikidata:Main_Page}}. These knowledge bases store structured, linked data that can be easily queried and visualised, making it easier for machines to understand the meaning of information on the Web.
Sentiment analysis \cite{kouadri2020quality} is another area of research in the Semantic Web for dealing with textual content. Determining the writer’s attitude towards a topic, whether positive, negative, or neutral, involves using natural language processing techniques. 

\paragraph{Modelling textual content.}
Modelling knowledge from textual content with the Semantic Web involves using ontologies to represent explicit formal, conceptual models and describe semantically unstructured content and databases \cite{rossiello2023knowgl}. This approach has already seen widespread adoption in the form of Schema.org and Linked Open Data (LOD) \footnote{\url{https://www.ontotext.com/blog/the-semantic-web-20-years-later/}}.

The process of modelling text content with ontologies involves analysing the collected text for a specific domain, identifying the relevant terms, concepts and their relationships, mapping and representing the ontology by representation language (e.g. OWL (Web Ontology Language), RDF (Resource Description Framework), or RDFS (Resource Description Framework Schema)), and finally evaluating the constructed ontology \cite{chimalakonda2020ontology}. The procedure of ontology construction can be done in one of three ways: manual construction, cooperative construction (need human intervention during the ontology constructing process) and (semi-) automatic construction \cite{al2020automatic}.

Considering the medical treatment of COVID-19, an ontology could be created by extracting key concepts and relationships from the text. For example, treatments, medications, and symptoms could be represented as concepts, and relationships such as “is used to treat,” “is a symptom of,” or “is a medication for” could be used to connect these concepts.
Here is an example of how an ontology for medical treatment on COVID-19 might look:
\begin{verbatim}
Concepts:   COVID-19, Antiviral Medications, 
            Nirmatrelvir with Ritonavir (Paxlovid), 
            Remdesivir (Veklury), Fever, 
            Cough, Shortness of Breath

Relationships:
Nirmatrelvir with Ritonavir (Paxlovid) is a medication for COVID-19
Remdesivir (Veklury) is a medication for COVID-19
Fever is a symptom of COVID-19
Cough is a symptom of COVID-19
Shortness of Breath is a symptom of COVID-19
\end{verbatim}

\paragraph{Reasoning (querying) with ontologies.}
Ontologies representing text can answer various queries about the domain they represent. One type of query that can be answered using ontologies representing text is conjunctive queries, the basic database queries \cite{bienvenu2020reasoning}. These are existentially quantified conjunctions of atoms, meaning they return all combinations of values that satisfy the conditions specified in the query. Conjunctive queries can retrieve information from an ontology by selecting the desired relationships between concepts.
Based on the ontology above, some example queries that could be used to retrieve information from it are:

\begin{verbatim}
What are the medications for COVID-19?
Query: 
SELECT ?x WHERE { ?x is a medication for COVID-19 }
Result: Nirmatrelvir with Ritonavir (Paxlovid), Remdesivir (Veklury)

What are the symptoms of COVID-19?
Query: SELECT ?x WHERE { ?x is a symptom of COVID-19 }
Result: Fever, Cough, Shortness of Breath

\end{verbatim}

The Clinical MetaData Ontology (CMDO) \footnote{\url{https://bmcmedinformdecismak.biomedcentral.com/articles/10.1186/s12911-019-0877-x}} is an example of an ontology representing clinical cases. It provides a simple classification scheme for elements of clinical data based on semantics and comprises 189 concepts under four first-level classes.

Reasoning on text ontologies involves deriving facts not explicitly expressed in the ontology or knowledge base. 
Reasoning with ontologies is becoming increasingly important in data-centric applications, where it helps to integrate heterogeneous data and provide a common conceptual vocabulary. 

\paragraph{Current trends and open issues.}
The following discusses the conclusions about how the combination of the Semantic Web, NLP, and AI (including ML and data mining) can be used to query textual content and extract knowledge from it. First, we address the main question of our analysis:
\begin{quote}
    {\em How can natural language processing (NLP) and the Semantic Web and machine learning/data mining contribute to model/query text content?}
\end{quote}

NLP and the Semantic Web can work together to model and query text content in several ways. For example, NLP techniques can be used to automatically extract meaningful information and semantics from text, which can then be represented using Semantic Web technologies such as RDF  and OWL. This allows for the creation of structured, linked data that can be easily queried and manipulated using Semantic Web query languages such as SPARQL \cite{maynard2017natural}.

In addition, NLP techniques such as named entity recognition, relation extraction, and sentiment analysis can be used to enhance the querying of text content by allowing for more sophisticated queries that take into account the meaning and context of the text. For example, a query could be constructed to find all documents that mention a specific person or organisation, or to find all documents that express a positive sentiment towards a particular topic \cite{maynard2017natural}.

\begin{enumerate}
    \item 
   {\em What are the barriers and challenges of textual content processing? What are representative applications? Which experiments  have been representative or  have posed particular challenges?}

  One of the main challenges is the difficulty of automatically extracting meaningful information and semantics from unstructured text. This process, known as ontology learning, involves using techniques from fields such as machine learning, text mining, knowledge representation and reasoning, information retrieval, and natural language processing to automatically acquire ontologies from unstructured text \cite{zaihrayeu2007web}.
Another challenge is the ambiguity and complexity of natural language. Natural language labels used to describe the contents of hierarchical classifications are easily understood by human users, but can be difficult for machines to reason about due to their ambiguity \cite{asim2018survey}. This creates a challenge for converting classifications into lightweight ontologies that can be used in the Semantic Web infrastructure.
In addition, building large ontologies is a difficult and time-consuming task, and it is not feasible to build ontologies for all available domains. This makes it necessary to develop automated methods for ontology learning that can scale to handle large amounts of data.  
    
   \item 	
   {\em How to deal with missing information, i.e., incomplete, poor quality or contradictory annotations on property graphs and/or ontologies representing text content? What are the impacts on querying and maintenance? What issues need to be addressed?}

   One approach to address this issue is to use quality assurance techniques to improve the quality of the ontology \cite{benbernou2017enhancing}. For example, one the study in \cite{mikroyannidi2014quality}  the detection of semantic regularities in entailed axioms can be used in ontology quality assurance, in combination with lexical techniques. Another approach is to use quality evaluation methodologies to assess the quality of the ontology and identify areas for improvement \cite{wilson2022ontology}. Additionally, there are tools and frameworks available that can help with the documentation and management of quality assurance measures for ontologies \cite{sheveleva2022ontology}.

Incomplete data in ontologies can have a significant impact on querying and maintenance. When data is incomplete, it can be difficult to accurately answer queries or make inferences based on the available information. This can lead to incorrect or incomplete results, which can affect the usefulness and reliability of the ontology.

To address these issues, several approaches have been developed. One approach is to use ontology-based data access (OBDA) techniques to access incomplete and/or heterogeneous data \cite{schneider2020ontologies}. These techniques allow for the use of ontologies to provide taxonomies and background knowledge that can help align and complete the data.
    
   \item 
   {\em Maintenance and updating of text content representations in the form of property graphs or ontologies:}
    \begin{itemize}
        \item How to make annotations evolve according to the evolution of repositories, ontologies and corpora?

         Ontology evolution techniques can detect changes in the ontologies and update the annotations accordingly. The work in \cite{cardoso2016leveraging} used annotators to generate more than 66 million annotations from a pre-selected set of 5000 biomedical journal articles and standard ontologies covering a period ranging from 2004 to 2016. The work highlighted the correlation between changes in the ontologies and changes in the annotations and discussed the necessity to improve existing annotation formalisms in order to include elements required to support (semi-) automatic annotation maintenance mechanisms.

Another approach is to use ontology alignment techniques to maintain the consistency of annotations across different ontologies \cite{pietranik2023methods,ziebelininteractive}.  It can be used to preserve the validity of ontology alignment using only the analysis of changes introduced to maintained ontologies. The precise definition of ontologies is provided, along with a definition of the ontology change log. A set of algorithms that allow revalidating ontology alignments have been built based on such elements \cite{pietranik2023methods}.

        \item How to validate new knowledge deduced, discovered or explicitly inserted?

        Validating new knowledge in an ontology can be done using a validation and verification driven ontology design process, such as the CQ-Driven Ontology Design Process (CODEP) proposed by Espinoza et al \cite{espinoza2021validation}. This process is driven by end-user requirements, defined as Competency Questions (CQs), and includes activities that validate and verify the incremental design of an ontology through metrics based on defined CQs. Another approach is to use a method to validate the insertion of a new concept in an ontology, such as the one presented by \cite{ngom2016method}. This method is based on finding the neighbourhood of the concept in the basic ontology and assessing its semantic similarity with its neighbourhood in a general ontology, then evaluating the correlation between the values found in these steps. There are also several tools available for ontology evaluation and validation, such as OntoQA \cite{tartir2010ontological}, which uses a set of metrics to measure different aspects of the ontology schema and knowledge base to give insight into the overall characteristics of the ontology. 
        
    \end{itemize}
\item 
{\em What are the possibilities of querying the different representations of the text content?}
\begin{itemize}

    \item How can we address querying in the "analytics" sense (data science queries with community discovery, centrality and link discovery operations) on graphs?
    
One way to query ontologies is through the use of SPARQL, a query language for RDF data. SPARQL allows users to write queries that can retrieve and manipulate data stored in RDF format. This can include data stored in triplestores, which are a type of database specifically designed for storing and querying RDF data.

Another approach to querying ontologies is through the use of reasoners, which can infer new knowledge from the existing data based on the logical rules defined in the ontology. This can allow for more complex queries that take into account the relationships between different entities and their properties.

In addition to these approaches, there are also various tools and frameworks available for querying ontologies like SPARQL, Protégé \footnote{\url{https://protege.stanford.edu/}} and Apache Jena \footnote{\url{https://learn.microsoft.com/en-us/azure/architecture/data-science-process/platforms-and-tools}}. They 
 can be used in conjunction with machine learning, but they do not have built-in machine learning functions.

   \item Is it relevant to think of a SPARQL-extended query language where the user can include a 'function' concerning graph data analysis? What are the advantages? What uses?  What challenges?

   SPARQL-extended query language where the user can include a ‘function’ concerning graph data analysis can be relevant. One example of such an extension is the addition of recursive features to the SPARQL query language \cite{della2011querying}, which allows for expressing analytical tasks over existing SPARQL infrastructure. This language is well-suited for both graph querying and analytical tasks and can express key analytical tasks on graphs \cite{hogan2020database}.

The advantages of such an extension include combining querying and analytics for graphs, allowing users to perform complex analytical tasks using a single query language. This can simplify the process of querying and analysing graph data making it easier for users to extract meaningful insights from their data \cite{hogan2020recursive}.

Some uses of this extended query language could include performing complex graph analytics, such as finding the shortest path between two nodes, computing centrality measures, or identifying clusters or communities within the graph \cite{mosser2018querying}.

One challenge in developing such an extension is ensuring it is compatible with existing SPARQL infrastructure and can be easily integrated into existing systems. Another challenge is ensuring that the language is expressive enough to support a wide range of analytical tasks while remaining easy to use and understand \cite{ali2022survey}.
\end{itemize}
\end{enumerate}

\section{Graph databases for storing, querying and analysing textual content}\label{sec:dbgraphs}
A graph database is a database management system (DBMS)  that uses graph structures, including nodes, edges, and properties, to represent and store data. The fundamental concept of a graph database is the graph, which relates data items in the store to a collection of nodes and edges, with the edges representing the relationships between the nodes. Graph databases prioritise relationships between data, making querying relationships fast and efficient. Relationships can also be intuitively visualised using graph databases, making them useful for heavily inter-connected data.

Graph databases are well adapted for managing data with a graph structure, for example, the Semantic Web, networks (social, transport, biology, etc.), and financial transactions.

\paragraph{Graph data models.}
There are two popular models of graph databases: property graphs and RDF graphs. Property graphs focus on analytics and querying, while RDF graphs emphasise data integration. Both graph models consist of a collection of points (vertices) and the connections between those points (edges).

A graph data model allows to represent data in a graph structure, where data points are represented as nodes and the relationships between them are represented as edges. Graph data modelling translates a conceptual view of data into a logical model. Therefore, decision-making must determine which entities in a dataset should be nodes, which should be relationships, and which should be discarded. The result is a blueprint of data’s entities, relationships, and properties.

In a graph database, nodes represent the fundamental data units, while edges represent the relationships between nodes. Properties are descriptive characteristics of nodes and edges that are not important enough to become nodes themselves. There is no formula for deriving a graph model from a dataset, but there are general guidelines that can help create an effective model. For example, it is essential to consider the types of questions  to answer with  data and design the graph database accordingly \footnote{\url{https://cambridge-intelligence.com/graph-data-modeling-101/}} \cite{prevoteau2022propagation}. 

 The textual content representation as graphs allows to perform traversal queries based on connections, use specific algorithms on graphs, find patterns, paths, communities, influencers, single points of failure, and other relationships and have a more intuitive understanding and visualisation of the data content.

\begin{figure*}[htbp]
\centerline{
\includegraphics[width=0.99\textwidth]{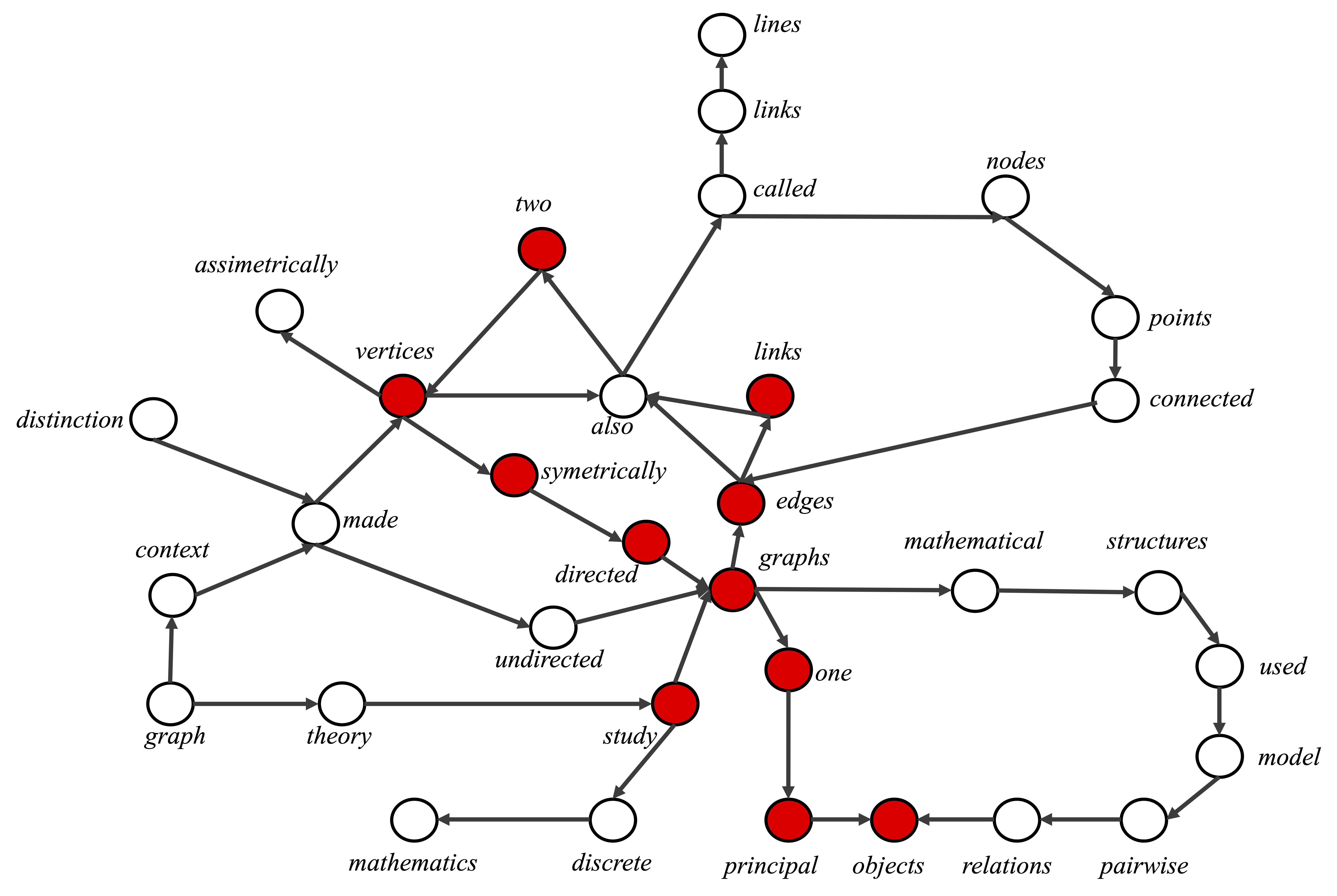}
}
\caption{Text graph example.}
\label{fig:dag-example}
\end{figure*}

\paragraph{Text graphs}
 represent a text item, such as a document, passage, or sentence, as a graph. Building a text graph is a preprocessing step in NLP to support text condensation, term disambiguation, topic-based text summarisation, relation extraction, and textual entailment. 
For creating a text graph, it is necessary first to identify the entities in the text that will be represented as nodes in the graph and then to identify the relationships between those entities that will be represented as edges between the nodes. The specific types of entities and relationships used will depend on the context and task. Entities can take various forms, such as individual words, bigrams, n-grams, or sequences of variable lengths. Relationships can represent a variety of connections between entities, such as adjacency in a sentence, co-occurrence within a fixed-length window, or some semantic or syntactic relationship. A  directed unweighted graph (see Figure \ref{fig:dag-example}) can be used for representing the following text (without stop words)  \footnote{This example was proposed in \url{https://towardsdatascience.com/keyphrase-extraction-with-graph-centrality-989e142ce427}}:

\begin{quote}
\footnotesize\it
    Centrality indices are answers to the question “What characterises an important vertex?” The answer is given in terms of a real-valued function on the vertices of a graph, where the values produced are expected to provide a ranking which identifies the most important nodes.The word “importance” has a wide number of meanings, leading to many different definitions of centrality.
\end{quote}

Text graphs are used in a variety of applications to represent and analyse textual data:
\begin{itemize}
    \item Information retrieval: Text graphs can represent the relationships between documents and the terms they contain, allowing for more effective information retrieval.
    \item Text summarisation: Text graphs can represent the relationships between sentences in a document, allowing for the automatic generation of summaries.
    \item Topic modelling: Text graphs can represent the relationships between documents and the topics they contain, automatically identifying topics within a corpus of text.
    \item Sentiment analysis: Text graphs can represent the relationships between words and their sentiment, allowing for automatic sentiment analysis in text.
\end{itemize}

\paragraph{Querying text graphs.}
Regular path queries (RPQs) are a way to navigate over paths in a graph using regular expressions \cite{10.14778/3489496.3489510}. A regular path query (RPQ) is a regular expression {\sf q} that returns all node pairs {\sf (u, v)} from a graph database that are connected by an arbitrary path labelled with a word from {\sf L(q)}. Intuitively, in an RPQ, the input is a graph, a source node and a regular expression. The goal is to identify all nodes connected to the source through a simple path whose label sequence satisfies the given regular expression. The regular expression is a formal specification of the search space that interests the user.
For example, one could use an RPQ to find all paths between two nodes that have a specific node in between or to find all nodes connected by paths of a certain form \footnote{\url{https://iccl.inf.tu-dresden.de/w/images/7/7f/Lecture-17-rpqs.pdf}}.

In addition to RPQ, other types of queries  can be used on text graphs:
\begin{itemize}
    \item Attributed graph queries extract information from attributed graphs, where vertices and edges are associated with attributes such as types, numbers, and texts. For example, suppose we have an attributed graph representing a document, where each node represents a word and has an attribute “word” containing the word itself. Edges represent the order of words in the document. We can use an RPQ to find all occurrences of a specific phrase in the document. If we want to find all occurrences of the phrase “artificial intelligence” in the document, we can use the following RPQ:
    \begin{verbatim}
        [word="artificial"]/./[word="intelligence"]
    \end{verbatim}
    \item Structure extraction is used to extract the underlying structure of a text graph by identifying and grouping equivalent nodes (see Schema extraction below). For example, For example, assuming that we have a graph as described in the previous example. Suppose we want to group all occurrences of the word “artificial” together. In that case, we can use structure extraction to identify all nodes representing the word “artificial” and group them into a single node. This would result in a simplified graph where all occurrences of the word “artificial” are represented by a single node. Similarly, we can use structure extraction to group nodes representing equivalent concepts, such as synonyms or related words. We could group nodes representing the words “artificial intelligence”, “AI”, and “machine learning” into a single node representing the concept of artificial intelligence.
\end{itemize}

\paragraph{Schema inference.}
Frequently, graph data comes without a predefined structure and in a constraint-less fashion, thus leading to inconsistency and poor quality. Inferring the schema can allow understanding the connections in the data better, maintaining its quality and exploiting graphs through navigational and AI-based queries. Extracting a schema from property graph instances can be challenging and complex, mainly when the instances exist before the schema definition \cite{lbath2021schema}. Neo4j provides a limited ability to view the schema of a database through a Cypher query that outputs a single-labelled directed graph. This graph displays the types of nodes that can be connected and the types of edges that can connect them. There are some limitations to viewing the schema of a database in Neo4j. For example, multi-labelled nodes in the graph instance are duplicated in the graph schema so that each node is assigned a single label. This results in the loss of label co-occurrence information, a critical property graph data model feature.
Additionally, properties, edge cardinality constraints, and node type hierarchies are not considered. GraphQL schemas can be derived from Neo4j databases using tools such as the Neo4j Desktop GraphQL plugin or neo4j-graphql-js. These tools can infer node and edge types and the data types of node properties. However, unlike other methods, they do not infer overlapping types, node hierarchies, or nested property values. Many approaches to schema inference involve identifying structural graph summaries by grouping equivalent nodes. Clustering techniques, such as those described in \cite{lutov2018statix} and \cite{bouhamoum2018scaling}, are commonly used to infer types in RDF datasets. These techniques are based on the assumption that the more properties two entities share, the more likely they are to belong to the same type. To achieve this, entities are grouped according to a similarity metric. Both techniques can handle hierarchical and overlapping types. In \cite{bouhamoum2018scaling}, a density-based clustering method called DBSCAN is used. In contrast, \cite{lutov2018statix} proposes a faster and more accurate clustering method called StaTIX, which uses the cosine similarity metric and is based on community detection. Property Graph (PG) schema inference in \cite{lbath2021schema} involves inferring types, basic and complex data types, overlapping types, and node hierarchies. 

\paragraph{Analysing text graphs.}
By converting free text into a graph representation, the implicit structure of the text is made explicit. This representation allows immediate access to information such as the most commonly used words (degree), the most frequently used n-grams, and the words most commonly used to convey information (paths in the graph between every two nodes), among other things. 

For example, extracting key phrases to return a list of relevant phrases (one or more words) from the input text can be possible. In this context, a relevant phrase is defined as one that is composed solely of candidate keywords, which are the most relevant unigrams. These candidate keywords are determined by graph centrality algorithms, which use the graph's structure to assign scores to nodes. In this case, the nodes represent words in the text and the relationships between them. Therefore, it is necessary to score the graph nodes (i.e., words) with a graph centrality algorithm \cite{arul2023graph,zhang2022pagerank}, extract candidate keywords (i.e., most relevant individual words), extract key phrases (based on the candidate keywords). 

In the previous example (see Figure \ref{fig:dag-example}, each node in the graph represents a term present in the input document (a single node for the multiple uses of the same term). Consider that after applying some centrality algorithm, the top 3 nodes are identified: (1) graphs: 0.56,
(2) study: 0.43, (3) objects: 0.35. 

For extracting the candidate keywords, we can consider a third of the total number of nodes as the number of candidate keywords. Note that the choice depends on the application context. The candidate keywords (ordered from highest to lowest scoring) are shown in figure \ref{fig:dag-example}: 
\begin{verbatim}
1. graphs
2. study 
3. objects
4. edges
5. vertices
6. two
7. link 
8. directed 
9. symmetrically 
10. principal 
11. one
\end{verbatim}

A key phrase in the input document is considered relevant and can be used for various tasks, including extractive summarisation. It is defined as a sequence of words in the input document composed solely of candidate keywords. For instance, if both “mathematical” and “structures” are candidate keywords, then “mathematical structures” would be considered a key phrase, and the words would not be considered individually. In the previous text and considering the candidate keywords, examples of key phrases can be: 
\begin{verbatim}
1. edges link two vertices symmetrically
2. directed graphs
3. edges link two vertices
4. principal objects
5. study
6. one
\end{verbatim}

\paragraph{Current trends and open issues.}
One of the main advantages of using a graph store compared to graph processing libraries  (e.g. NetworkX) is that graph stores are designed to handle large amounts of data and provide persistence, ACID properties, CRUD operations, commits, indexing, and other database features. This means it is possible to run graph algorithms as often as required without recreating the graph each time.

\begin{quote}
    {\em How can natural language processing, the Semantic Web, DB design for Graph DBMSs, and machine learning/data mining contribute to model/query text content?}
\end{quote}
Graph databases are focused on efficiently storing and querying highly connected data. They can use various data models for graphs and their data extensions. Some examples of graph databases include labelled property graphs \cite{pokorny2018functional}, RDF graphs \cite{ali2022survey}, and hypergraphs \cite{bellaachia2015short}.
On the other hand, an ontology is a formal and standardised representation of knowledge used to break down data silos. It allows for defining concepts and relationships between them, emphasising class inheritance so subject matter experts can appropriately label the data in their databases.  
Thus, modelling text content with labelled property graphs involves representing the data as a set of nodes and edges with properties, while modelling text content with ontologies consists of defining concepts and relationships between them formally and standardised. 

\begin{enumerate}
    \item 
   {\em What are the barriers and challenges of textual content processing? What are representative applications? Which experiments  have been representative or  have posed particular challenges?}

    The challenges and barriers associated with textual content processing with graph databases include scalability, partitioning, processing complexity, and hardware configurations. For example, scaling graph data by distributing it in a network is much more complex than scaling simpler data models and is still a work in progress \cite{pokorny2015graph}. Mining complete frequent patterns from graph databases is also challenging since supporting operations are computationally costly. Also, partitioning and processing graph-structured data in parallel and distributed systems is challenging.
    
    
\item 
{\em What are the possibilities of querying the different representations of the text content?}

   Text content can be represented as different types of graphs, and every kind of graph has specific methods for querying and retrieving information. For instance, if the text content is represented as a knowledge graph, one could use a query language such as SPARQL to retrieve information from the graph. Similarly, if the text content is represented as an RDF graph, one could use SPARQL or another RDF query language to retrieve information from the graph. If the text content is represented as a property graph, one could use a query language like Cypher or Gremlin to retrieve information from the graph. 
   
    
    
    \item {\em What are the solutions proposed by graph-based DBMSs? How can they be positioned concerning solutions such as ML pipelines/DS pipelines?}

Many graph databases, including Amazon Neptune, OrientDB, ArangoDB, JanusGraph, and TigerGraph, can analyse graphs using machine learning algorithms \cite{zhang2021automated}. Graph-based DBMSs offer a variety of analytics and data science solutions for processing graphs. These solutions include using platforms for enterprise knowledge graphs or graph artificial intelligence (AI) and graph-based deep learning approaches to improve analytics. The graph DBMS market vendors are expanding their stacks into platforms for enterprise knowledge graphs or graph AI with associated product ecosystems. For example, graph DBMS can use a graph convolution network (GCN) to enable CRM analytics to forecast sales.

The Neo4j Graph Data Science addon provides a powerful set of tools for analysing relationships in data using graph algorithms and machine learning. The library includes a catalogue of 65+ graph algorithms, graph native ML pipelines, and graph visualisation tools, allowing users to answer questions such as what is important, what is unusual, and what is next.

Amazon Neptune has a feature called Neptune ML that uses Graph Neural Networks (GNNs) to make predictions using graph data. Neptune ML can improve the accuracy of most predictions for graphs by over 50\% when compared to making predictions using non-graph methods \footnote{\url{https://aws.amazon.com/neptune/machine-learning/}}.

OrientDB has a suggestion algorithm that leverages graph-based data modelling and machine-learning techniques to generate intelligent recommendations \footnote{\url{https://saturncloud.io/blog/orientdb-suggestion-algorithm-enhancing-data-analysis-with-intelligent-recommendations/}}.

ArangoDB has a Graph Data Science Library that includes over 50 algorithms for dependencies, clustering, similarity, matching/patterns, flow, centrality, and search \footnote{\url{https://www.arangodb.com/graph-analytics-at-enterprise-scale/}}. Many machine learning algorithms are based on graphs.

JanusGraph supports graph processing. Scaling graph data processing for real-time traversals and analytical queries is JanusGraph’s foundational benefit \footnote{\url{https://docs.janusgraph.org/}}.

TigerGraph has an in-database graph data science library that includes over 50 algorithms for dependencies, clustering, similarity, matching/patterns, flow, centrality, and search \footnote{\url{https://www.tigergraph.com/graph-data-science-library/}}.

\end{enumerate}

\section{Beyond facts, analysing and discovering knowledge from text graphs}\label{sec:ml}
 After extracting information from text, a significant amount of work is still required to transform that information into knowledge and valuable insights. These insights may take the form of discoveries or the confirmation and verification of previous hypotheses by creating new connections within our existing knowledge of a particular domain. Artificial intelligence applied to graphs is used for this purpose.



 \paragraph{Graph analytics}
 or Graph algorithms are analytic tools used to determine the strength and direction of relationships between objects in a graph. For example, detecting cybercrimes such as money laundering, identity fraud and cyberterrorism. Applying graph analysis to social networks and communities, such as monitoring statistics, identifying influencers and analysing the traffic and quality of service for computer networks.

 Graph data can be ingested into machine learning algorithms and then be used to perform classification, clustering, regression, etc. Together, graph and machine learning provide greater analytical accuracy and faster insights. For example, graph features can work as an input for machine learning, like using PageRank score as a feature in a machine learning model. 
Machine learning can also enhance graph, for example, performing entity resolution \footnote{
Entity resolution is the process that resolves entities and detects relationships. The pipelines perform entity resolution as they process incoming identity records in three phases: recognise, resolve, and relate.} and combine different data sources into one single graph \footnote{\url{https://www.ibm.com/docs/en/iii/9.0.0?topic=insight-entity-resolution} }\cite{getoor2012entity,christophides2020overview}. 

A graph database and a graph data science platform can be utilized to calculate various graph metrics, including generic metrics such as betweenness centrality \cite{grando2018machine} and PageRank \cite{sargolzaei2010pagerank,zhang2022pagerank} score, as well as domain-specific metrics such as the number of known fraudsters indirectly connected to a given client. Additionally, these tools can be used to generate graph embeddings \cite{wang2022survey}, which translate graph data into a more suitable format for machine learning.

 \paragraph{Machine learning with Graphs}
  is a powerful tool for addressing a wide range of problems and gain new insights into a wide range of domains. Some typical problems that can be addressed with machine learning and graphs include recommendation systems, community detection, link prediction, and node classification. Graphs can be used to represent the relationships between users and items, and machine learning algorithms can be applied to these graphs to: 
 \begin{itemize}
     \item make personalised recommendations;
     \item identify clusters or communities of nodes that are more densely connected to each other than to the rest of the graph;
     \item  predict the likelihood of a link forming between two nodes in a graph, based on the characteristics of the nodes and their relationships with other nodes in the graph;
     \item classify the nodes in a graph based on their attributes and their relationships with other nodes in the graph. 
 \end{itemize}

\begin{figure*}[htbp]
\centerline{
\includegraphics[width=0.99\textwidth]{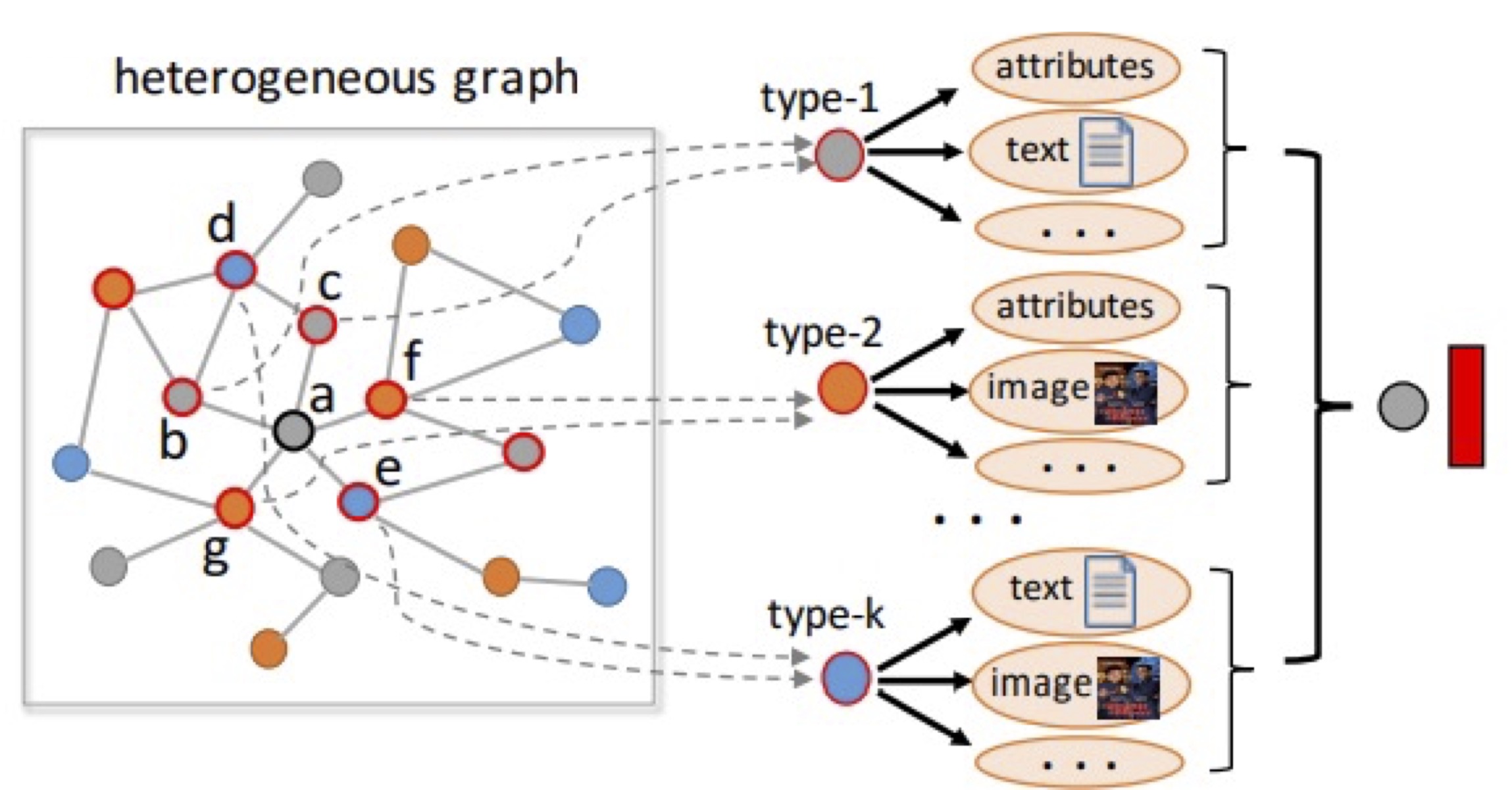}
}
\caption{Heterogeneous graph.}
\label{fig:heterogeneous-graph}
\end{figure*}

Graph Neural Networks (GNNs) are a type of neural network designed to work with graph data structures.  
GNNs are used in predicting nodes, edges, and graph-based tasks.
The key design element of GNNs is the use of pairwise message passing, such that graph nodes iteratively update their representations by exchanging information with their neighbours. This method allows dealing with dynamic graphs \cite{liu2020detecting}. GNN can also allow dealing with heterogeneous data represented as heterogeneous graphs \cite{zhang2019heterogeneous}. The principle is to create view from an heterogeous graph\footnote{A heterogeneous graph contains either multiple types of objects or multiple types of links. It can be used to model complex systems and relational data.}  that contains nodes and edges of the same type and look for a vectorial representation for each type of node as shown in figure \ref{fig:heterogeneous-graph}.

Vectorial representations or embeddings capture the graph topology, vertex-to-vertex relationship, and other relevant information about graphs, subgraphs, and vertices. These representations can encode each vertex with its vector representation to perform visualisation or prediction on the vertex level. For example, vertices can be visualised in a 2D plane, or new connections can be predicted based on vertex similarities. 

Graph embeddings, conversely, are the transformation of property graphs to a vector or a set of vectors. These embeddings can be used to make predictions on the graph level and to compare or visualise whole graphs. For instance, graph embeddings can be used to compare chemical structures. For example, suppose we have two graphs representing social networks, and we want to compare the similarity of their community structures. We can use a graph embedding algorithm to generate embeddings for each node in both graphs. Then, we can use a similarity measure, such as cosine similarity, to compare the embeddings of corresponding nodes in the two graphs. This will give us a measure of how similar the community structures of the two graphs are.

Overall, vectorial representations and embeddings are powerful tools for analyzing and understanding complex graph data.
An essential challenge is a decision on the embedding dimensionality. Longer embeddings preserve more information and induce higher time and space complexity than sorter embeddings. Users need to make a trade-off based on the requirements \footnote{\url{https://towardsdatascience.com/graph-embeddings-the-summary-cc6075aba007}}. 

Relevant application domains for GNNs include Natural Language Processing, social networks, citation networks, molecular biology, chemistry, physics and NP-hard combinatorial optimisation problems. Several open-source libraries implementing graph neural networks are available, such as PyTorch Geometric (PyTorch), TensorFlow GNN (TensorFlow), and jraph (Google JAX).

 \paragraph{Data Science Pipelines on Text graphs.}
The data science pipeline involves a series of steps to gather raw data from multiple sources, analyse it, and present the results in an understandable format. This process can be applied to constructing, exploring, and analysing text graphs. General templates for these tasks can be adapted based on the algorithms and strategies used. For example, a graph construction pipeline from textual input may use information retrieval strategies to create a raw graph with relevant terms from the document represented as nodes. Rules can be established to connect the nodes based on context, such as connecting terms representing characters if separated by no more than 5 words in the text. The raw graph can then be pruned to remove dead nodes, duplicates, and isolated nodes. 

\begin{figure*}[htbp]
\centerline{
\includegraphics[width=0.99\textwidth]{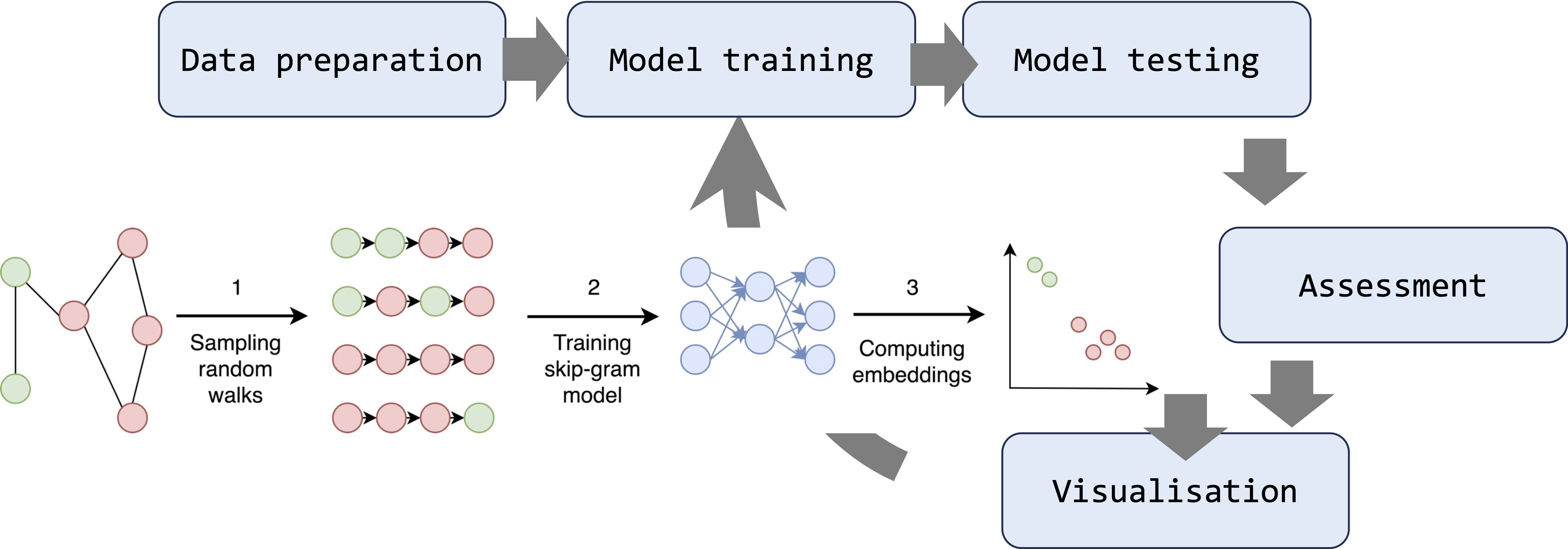}
}
\caption{Graph processing pipeline example.}
\label{fig:pipeline-example}
\end{figure*}

To preprocess the text graph and extract knowledge from texts, the pipelines should include tasks for transforming the graphs into vectorial representations that can be used in training, testing, and evaluating GNN models. Depending on the type of graph (i.e., heterogeneous or homogeneous graph, directed or undirected, properties in the nodes and/or edges), the preprocessing pipeline can change. For example, extracting nodes and edges separately with their properties and transforming them to vectors using text2vec. In contrast, selecting graph views from a heterogeneous graph, such that the nodes and edges are the same type, and then computing a global vector representing the whole graph. Calibrating the transformation models is a recurrent process that data science pipelines can exhibit (see Figure \ref{fig:pipeline-example} \footnote{\url{https://towardsdatascience.com/graph-embeddings-the-summary-cc6075aba007}}).

 \paragraph{Current trends and open issues.}
 \begin{quote}
    {\em How can 
    data science on graphs
    contribute to model/query text content?}
\end{quote}
Data science on graphs can contribute to modelling and querying text content by providing a powerful way to represent and analyse the relationships between entities in a dataset. This can improve accuracy and performance for text classification, information retrieval, and natural language processing tasks. 

\begin{enumerate}
    \item 
    {\em What are the barriers and challenges of textual content processing? What are representative applications? Which experiments  have been representative or  have posed particular challenges?}\\
     Processing natural language involves enabling programs to produce relevant documents in good shape and automatically marking and extracting relevant information buried in the text. Preparing textual data and moving it to a target system for analytics (textual Extraction Transformation and Loading - ETL) involves identifying patterns and trends in the data.
Standard natural language processing approaches gain valuable knowledge in business process complexity, but this is still an active area of research. An example of a challenging application is processing the textual content in clinical cases \cite{agrawal2020medical}: identifying important clinical terms stated in digital text, such as treatments, disorders, and symptoms, can be challenging; locating key entities within the text: clinical text may be in an unstructured format, making it difficult to process; repeated words make it difficult to extract meaningful information.

    \item 	
    {\em How to deal with missing information, i.e., incomplete, poor quality or contradictory annotations on property graphs and/or ontologies representing text content? What are the impacts on querying and maintenance? What issues need to be addressed?}\\
    Dealing with missing information in text graph databases can be done  using gradual pattern extraction techniques \cite{shah2020handling} and building an information extraction pipeline \footnote{\url{https://neo4j.com/blog/text-to-knowledge-graph-information-extraction-pipeline/}}.  The pipeline involves taking unstructured text inputs, processing them with natural language processing (NLP) techniques, and using the resulting structures to populate or enrich a knowledge graph. This can help to fill in missing values and improve the overall quality of the data.
    
    \item {\em Maintenance and updating of text content representations in the form of property graphs or ontologies.}\\
    The approaches to making annotations evolve according to the evolution of repositories, including using an information extraction pipeline or gradual pattern extraction techniques \cite{cardoso2016leveraging}. In the first case, unstructured text inputs are processed with natural language processing (NLP) techniques, and the resulting structures populate or enrich a knowledge graph. In the second case, extracting gradual patterns from property graphs provides end-users with tools for mining correlations in the data when missing values exist. 
\item 
{\em What are the possibilities of querying the different representations of the text content?
   %
    How can we address querying in the "analytics" sense (data science queries with community discovery, centrality and link discovery operations) on graphs?} \\
   The method used to query graphs depends on the type of representation used and the user's specific needs. For instance, vector similarity search is a technique that searches for vectors based on their similarity to a given query\footnote{A vector search database, also known as a vector similarity search engine or vector database, is a type of database that is designed to store, retrieve, and search for vectors based on their similarity given a query \url{https://labelbox.com/blog/how-vector-similarity-search-works/}.}. This approach is commonly used in applications such as image retrieval, natural language processing, and recommendation systems. Knowledge graphs, on the other hand, integrate heterogeneous data and allow for querying large amounts of data using query languages such as SPARQL \cite{liang2021querying}.

Several visual representations of text content can be queried. Concept maps are diagrams that visually represent relationships between concepts and ideas and can be used to uncover the logical structure of arguments and identify unstated assumptions. Mind maps are diagrams that visually organise information around a central idea or topic and can be searched for specific content using search and filter operations. Argument maps are visual representations of the structure of an argument, including all the key components, such as the conclusion and premises.
    
\end{enumerate}
\section{Conclusions and Outlook}\label{sec:conclusion}
Graph technology forms the foundation of modern data and analytics, with capabilities to enhance and improve user collaboration, machine learning models, and explainable AI. 
 The  disciplines addressing text processing are guided by different objectives: 
\begin{enumerate}
\item Content extraction: This involves using linguistic techniques to delve into the language and extract meaningful content.
\item Knowledge representation: This assumes that textual content defines a network of representative concepts, semantic relations, and associated consistency rules that represent the knowledge contained in the text, including knowledge that can be inferred from it.
\item Efficient query execution: This involves selecting appropriate graph data models to represent textual content so that specific queries can be answered efficiently.
\item Knowledge modelling, discovery, and prediction: Given textual content represented as graphs, this involves modelling how concepts weave content by connecting words, sentences, and groups of sentences, with the hypothesis that new knowledge can be produced and predicted. For example, graph machine learning, also known as geometric machine learning, can learn from complex data such as graphs and multi-dimensional points. Its applications have been relevant in fields such as biochemistry, drug design, and structural biology\footnote{\url{https://www.gartner.com/smarterwithgartner/gartner-top-10-data-and-analytics-trends-for-2021}}. 
\end{enumerate}

The objective of analysing textual data is not only to extract and infer knowledge using artificial intelligence techniques such as machine learning, knowledge representation, and reasoning, but also to learn from the textual content and generate new knowledge. This challenge requires the integration of results to achieve previous goals. The vision is to preserve and compare the knowledge extracted from textual content, represented by graphs, within various consistency spaces. The challenge is to convey knowledge from textual content as a dynamic entity, which has already been accomplished using ontologies and graph databases, and to model its evolution and how it can be evaluated and validated within a learning process. New developments in large language models, generative AI, and reinforcement learning are propelling this ambitious trend forward. The challenges introduced necessitate the creation of new data management paradigms that take into account the characteristics of data encoding requirements (e.g., vectors and matrices) and use these encodings as first-class citizens for efficient processing. Vectorial and neural data management systems will provide persistence, indexing, and maintenance support to enable learning models to operate at scale without impedance overhead.

\section*{Acknowledgements}
The paper summarises the discussion in the panel of the national action DOING in the national symposium MADICS 2022 in Lyon. We thank the panellists who shared their insight and perspectives about the questions addressed in this paper: Donatello Conte (Polytech Tours, LIFAT),
Nathalie Hernandez (Université de Toulouse, IRIT),
Catherine Roussey (INRAE),
Agata Savary (Université Paris-Saclay, LISN),
Nicolas Travers (ELSIV, Centre de Recherche Da Vinci).

%
%
 \bibliographystyle{splncs04}
 \bibliography{mybibliography}

\end{document}